\documentclass[10pt]{article}
\usepackage{epsf}

\begin{document}
\title{Electronic Detection of Gravitational
Disturbances and Collective Coulomb Interactions}
\author{A. Widom, D. Drosdoff, S. Sivasubramanian, Y.N. Srivastava$^\dagger $,  \\
Physics Department, Northeastern University, Boston MA USA \\
$^\dagger$ Physics Department \& INFN, University of Perugia, Perugia Italy}
\date{}
\maketitle
\begin{abstract}
The cross section for a gravitational wave antenna to absorb a graviton
may be directly expressed in terms of the non-local viscous response function
of the metallic crystal. Crystal viscosity is dominated by electronic processes
which then also dominate the graviton absorption rate. To compute this rate
from a microscopic Hamiltonian, one must include the full Coulomb interaction
in the Maxwell electric field pressure and also allow for
strongly non-adiabatic transitions in the electronic kinetic pressure.
The view that the electrons and phonons constitute ideal gases with a weak
electron phonon interaction is not sufficiently accurate for estimating the
full strength of the electronic interaction with a gravitational wave.
\end{abstract}

\section{Introduction \label{CM}}

Resonant acoustic modes in massive metallic bars have long been used as
a probe for detecting possible gravitational wave sources. The interaction
of gravitational waves with such antennae was thought by many to be
dominated by the heavy nuclear masses within the metal. The interaction
with the lighter electron masses was considered to be negligible. This
view is valid only for static Newtonian gravity, which
couples to the mass density \begin{math} \rho  \end{math}. We have recently
shown\cite{us} that gravitational waves couple into the pressure tensor
\begin{math} {\sf P} \end{math}, and the pressure is dominated by electronic motions.
Thus, the electronic coupling to the gravitational wave cannot be ignored.

The condensed matter Hamiltonian required to understand the coupling of the
electrons to the gravitational wave is the sum of kinetic energies of the electrons
and the nuclei plus the Coulomb interactions between all of the charged particles.
Some standard models of the solid state which employ free electron and free phonon
gases together with a weak electron-phonon interaction are inadequate\cite{them} for
the description of electron-graviton interactions since terms leading to highly
excited virtual electronic states have been incorrectly thrown away. For example,
the electron-electron Coulomb interactions\cite{elel} are not considered in the
usual electron-phonon interaction model. Model Hamiltonians are considered in
Sec.\ref{CM}. From the Hamiltonian one can compute the pressure tensor which
determines the interaction between the gravity wave and the antenna. The exact
result is then the kinetic pressure plus the Maxwell field pressure. The Coulomb terms
in the Maxwell stress tensor are crucially important for understanding the mutual
interactions between electrons, phonons and gravitons as discussed in Sec.\ref{GI}.
In Sec.(\ref{absorb}) the rigorous expression for the single graviton absorption
cross section is exhibited and shown to be expressed directly and rigorously
in terms of the non-local viscosity of the crystal. Since the electrons dominate
the viscosity, they also dominate the absorption cross section.

\section{Condensed Matter\label{CM}}

The metal bar may be described by a Hamiltonian \begin{math} {\cal H} \end{math}
which consists of the kinetic energy \begin{math} {\cal K}_n \end{math} of the
nuclei and the kinetic energy \begin{math} {\cal K}_{el} \end{math} of the electrons
all interacting with the Coulomb law \begin{math} {\cal U} \end{math},
\begin{eqnarray}
{\cal H}&=& {\cal K}+{\cal U},
\nonumber \\
{\cal K} &=& {\cal K}_{el}+{\cal K}_n=
-\sum_j \left(\frac{\hbar^2}{2m}\right)\nabla_j^2-
\sum_a\left(\frac{\hbar^2}{2M_a}\right) \nabla_a^2,
\nonumber \\
{\cal U}&=& e^2\left\{\sum_{i<j}\frac{1}{r_{ij}}+
\sum_{a<b}\frac{Z_aZ_b}{R_{ab}}-
\sum_{i,a}\frac{Z_a}{|{\bf r}_i-{\bf R}_a|} \right\}.
\label{CoulombUs}
\end{eqnarray}
The pressure implied by Eq.(\ref{CoulombUs}) is dominated by the
electronic motions rather than the nuclear motions so that the
electronic coupling to the gravitational wave becomes crucial
for determining the detection efficiency. We have found that the
efficiency induced by including the electronic coupling to gravity is
considerably enhanced above the efficiency found by including
only the nuclear coupling to gravity.

The considerations above have been recently criticized\cite{them}.
The microscopic Hamiltonian employed\cite{them} for the crystal is a standard
electron-phonon {\em approximation}\cite{abgd} to our Hamiltonian Eq.(\ref{CoulombUs})
given (in first quantized notation) as
\begin{eqnarray}
{\cal H}_{eff}&=& {\cal K}_{el}+{\cal H}_{ph}+{\cal H}_{el-ph},
\nonumber \\
{\cal K}_{el}&=&-\left(\frac{\hbar^2}{2m}\right)\sum_j \nabla_j^2,
\nonumber \\
{\cal H}_{ph}&=&-\frac{\hbar^2 }{2}
\sum_k \left(\frac{\partial }{\partial Q_k}\right)^2
+\frac{1}{2}\sum_k \omega_k^2 Q_k^2,
\nonumber \\
{\cal H}_{el-ph}&=& \sum_j \Phi ({\bf r}_j)
\ \ \ \ {\rm where}\ \ \ \ \Phi({\bf r})=\sum_k Q_k \phi_k({\bf r}),
\label{ApproxThem}
\end{eqnarray}
or in second quantized notation
\begin{equation}
{\cal H}_{eff} =\int \psi^\dagger ({\bf r})
\left\{-\frac{\hbar^2}{2m}\nabla^2+\Phi({\bf r})\right\}
\psi({\bf r})d^3{\bf r}+
\sum_k \left(b_k^\dagger b_k+\frac{1}{2}\right)\hbar \omega_k .
\label{ApproxThem2nd}
\end{equation}

Starting from the {\em exact} Coulomb Hamiltonian
\begin{math} {\cal H} \end{math}
one may derive the low energy effective Hamiltonian
\begin{math} {\cal H}_{eff} \end{math}
only by employing a sequence of approximations\cite{el_ph}:
(i) The electron-electron Coulomb interactions are ignored.
(ii) Phonon modes are derived in the adiabatic approximation.
(iii) The non-adiabatic excitation interaction matrix Hamiltonian
is replaced by a local electron deformation potential
\begin{math} \Phi ({\bf r}) \end{math}. These three approximations make
the effective Hamiltonian inadequate for discussing the gravitational
wave interaction. In our work\cite{us} we avoided these three
approximations by employing rigorously exact sum rules.
Any other work\cite{them} which starts from the antenna
Hamiltonian \begin{math} {\cal H}_{eff} \end{math} is bound to miss
the electronic-gravitational enhanced efficiency because of
an inadequate approximation in the  Hamiltonian
from which the computation begins.

\section{Pressure and Gravitational Interactions\label{GI}}

For a small gravitational wave strain
\begin{math} {\sf u}({\bf r},t) \end{math} described by
the space-time metric
\begin{equation}
c^2 d\tau^2 = c^2dt^2-|d{\bf r}|^2
-2d{\bf r}\cdot {\sf u}({\bf r},t)\cdot d{\bf r},
\label{metric}
\end{equation}
the interaction between the gravitational wave and condensed
matter is described by
\begin{equation}
{\cal H}_{int}=-\int ({\sf P:u})d^3{\bf r},
\label{gravint}
\end{equation}
wherein \begin{math} {\sf P} \end{math} is the pressure tensor of
the condensed matter. For the exact Coulomb Hamiltonian the pressure tensor
is given by the sum of the kinetic pressure and the
Maxwell field pressure\cite{Pauli}
\begin{equation}
{\sf P}({\bf r})={\sf P}_{\cal K}({\bf r})+{\sf P}_{Maxwell}({\bf r}).
\label{pressure1}
\end{equation}
The electron and nuclear momenta will be denoted, respectively by
\begin{math} {\bf p}_i=-i\hbar \nabla_i \end{math} and
\begin{math} {\bf P}_a=-i\hbar \nabla_a \end{math}. Spatial indices
will be denoted by \begin{math} \mu {\rm \ and\ }\nu  \end{math} which may
be \begin{math} x,y {\rm \ or\ } z \end{math}. The kinetic pressure tensor
\begin{eqnarray}
{\sf P}_{\cal K}(\bf r) &=& {\sf P}_{{\cal K}_{el}}({\bf r})
+{\sf P}_{{\cal K}_{n}}({\bf r}),
\nonumber \\
{\sf P}_{{\cal K}_{el}}({\bf r})_{\mu \nu}&=&\frac{1}{4m}\sum_j
\left({\bf p}_{j\mu}{\bf p}_{j\nu}\delta ({\bf r}-{\bf r}_j)+
{\bf p}_{j\mu}\delta ({\bf r}-{\bf r}_j){\bf p}_{j\nu}\right)
\nonumber \\
&+&\frac{1}{4m}\sum_j\left({\bf p}_{j\nu }\delta ({\bf r}-{\bf r}_j){\bf p}_{j\mu }
+\delta ({\bf r}-{\bf r}_j){\bf p}_{j\mu }{\bf p}_{j\nu}\right),
\nonumber \\
{\sf P}_{{\cal K}_{n}}({\bf r})_{\mu \nu}&=&\sum_a \frac{1}{4M_a}
\left({\bf P}_{a\mu}{\bf P}_{a\nu}\delta ({\bf r}-{\bf R}_a)+
{\bf P}_{a\mu}\delta ({\bf r}-{\bf R}_a){\bf P}_{a\nu}\right)
\nonumber \\
&+&\sum_a
\frac{1}{4M_a}\left({\bf P}_{a\nu }\delta ({\bf r}-{\bf R}_a){\bf P}_{a\mu }
+\delta ({\bf r}-{\bf R}_a){\bf P}_{a\mu }{\bf P}_{a\nu}\right),
\label{pressure2}
\end{eqnarray}
and the Maxwell field pressure
\begin{eqnarray}
{\sf P}_{Maxwell}({\bf r})&=&\frac{1}{8\pi }
\sum_{(i,j\ne )}\left[{\bf E}_i({\bf r})\cdot {\bf E}_j({\bf r}){\sf 1}
-2{\bf E}_i({\bf r}){\bf E}_j({\bf r})\right]
\nonumber \\
&+&\frac{1}{8\pi }
\sum_{(a,b\ne )}\left[{\bf E}_a({\bf r})\cdot {\bf E}_b({\bf r}){\sf 1}
-2{\bf E}_a({\bf r}){\bf E}_b({\bf r})\right]
\nonumber \\
&+&\frac{1}{8\pi }
\sum_{(a,i)}\left[{\bf E}_a({\bf r})\cdot {\bf E}_i({\bf r}){\sf 1}
-{\bf E}_a({\bf r}){\bf E}_i({\bf r})-{\bf E}_i({\bf r}){\bf E}_a({\bf r})\right],
\label{pressure3}
\end{eqnarray}
wherein the electric fields due to an electron or a nucleus are given,
respectively, by
\begin{eqnarray}
{\bf E}_i({\bf r}) &=& -\frac{e({\bf r}-{\bf r}_i)}{|{\bf r}-{\bf r}_i|^3},
\nonumber \\
{\bf E}_a({\bf r}) &=& \frac{eZ_a({\bf r}-{\bf R}_a)}{|{\bf r}-{\bf R}_a|^3}.
\label{electric1}
\end{eqnarray}

Employing the {\em approximate} Hamiltonian in Eq.(\ref{ApproxThem}),
Branchina et. al.\cite{them} imply a pressure (via the linear
Hamiltonian coupling in \begin{math} h_{\mu \nu}=2u_{\mu \nu} \end{math})
which is much simpler than the rigorous
Eqs.(\ref{pressure1}), (\ref{pressure2}) and (\ref{pressure3}) of our
work; i.e.
\begin{equation}
{\sf P}_{eff}({\bf r})\approx {\sf P}_{{\cal K}_{el}}({\bf r})
+{\sf P}_{ion}({\bf r}).
\label{pressure4}
\end{equation}
The simplicity of the above approximate
\begin{math} {\sf P}_{eff}({\bf r}) \end{math} compared
with the exact \begin{math} {\sf P}({\bf r}) \end{math}
arises because so very many of the terms in the Maxwell pressure tensor
Eq.(\ref{pressure3}) have been simply thrown away.
For example, all Coulomb electric field interactions
between electrons have been ignored.

\begin{figure}[tp]
\epsfxsize=7.0cm\centerline{\epsffile{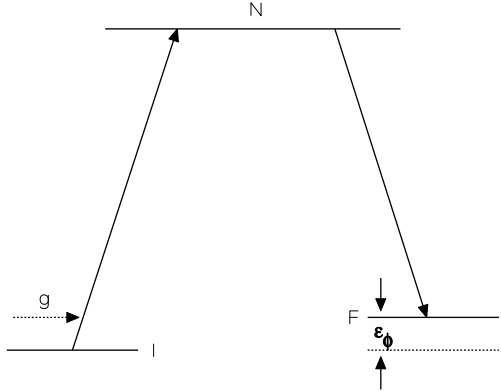}}
\caption{An incident graviton of energy $\hbar \omega_g$ excites the antenna
from an initial state $I$ to a final state $F$ containing a phonon with
energy $\varepsilon_\phi $. The reaction requires the virtual electron-hole pair
$N$ present only if the full pressure ${\sf P}$ is used to compute the matrix
element $\left<F\right|{\cal H}_{int}\left|g,I\right>
=-\int \left<F\right|{\sf P}\left|I\right>:{\sf u}_g d^3{\bf r}$
for the process.}
\label{FIG1}
\end{figure}
To see what is involved, consider a process in which a graviton
\begin{math} g \end{math} with energy \begin{math} \hbar \omega_g \end{math}
is absorbed by an antenna producing a phonon \begin{math} \phi \end{math}
with energy \begin{math} \epsilon_\phi \end{math}; i.e.
\begin{equation}
g+I\to F\ \ \ {\rm where}\ \ \ \hbar \omega_g=\epsilon_\phi
\label{g_scatter1}
\end{equation}
With the exact Coulomb Pressure of Eqs.(\ref{pressure1})-(\ref{pressure3}),
but {\em not} with the approximate pressure of Eq.(\ref{pressure4}),
an electronic process can occur with an intermediate electron-hole pair
state \begin{math} N \end{math}
\begin{equation}
g+I\to N\to  F
\label{g_scatter2}
\end{equation}
as shown in Fig.(\ref{FIG1}).

For example, with the help of the full Maxwell stress tensor one computes
the pressure matrix elements including highly excited virtual
non-adiabatic excited states,
\begin{equation}
\left<F\right|{\sf P}\left|I\right>=\frac{1}{8\pi }\sum_N
\left\{\left<F\right|{\bf E}\left|N\right>\cdot
\left<N\right|{\bf E}\left|I\right>{\bf 1}-2
\left<F\right|{\bf E}\left|N\right>
\left<N\right|{\bf E}\left|I\right>\right\}+\ldots ,
\label{pressurematrix}
\end{equation}
in which the electron pressure
\begin{math} {\sf P}_{{\cal K}_{el}} \end{math}
included in (\begin{math} \ldots  \end{math})
also has a similar structure for the intermediate states.

\section{Viscosity \label{absorb}}

In an infinite medium, the linear approximation to the Einstein field
equation is
\begin{equation}
\left\{\frac{1}{c^2}\left(\frac{\partial }{\partial t}\right)^2
-\nabla^2 \right\}{\sf u}=\left(\frac{8\pi G}{c^4}\right){\sf p}.
\label{absorb1}
\end{equation}
The transverse traceless part \begin{math} {\sf p} \end{math} of the pressure
\begin{math} {\sf P} \end{math} is related to the strain
\begin{math} {\sf u} \end{math} by the constitutive relation
\begin{equation}
{\sf p}=-2\left(\mu {\sf u}+
\eta \frac{\partial {\sf u}}{\partial t}\right),
\label{absorb2}
\end{equation}
wherein \begin{math} \mu  \end{math} is a Lam\'e elastic constant
and \begin{math} \eta  \end{math} is the crystal viscosity.
The gravitational wave then travels through the elastic media
as described by the wave equation
\begin{equation}
\left\{\frac{1}{c^2}\left(\frac{\partial }{\partial t}\right)^2
+\frac{16\pi G\eta }{c^4}\left(\frac{\partial }{\partial t}\right)
+\frac{16\pi G\mu }{c^4}-\Delta \right\}{\sf u}=0.
\label{absorb3}
\end{equation}
The energy in the wave attenuates at rate \begin{math} \Gamma \end{math}
(per unit time) as determined by Eq.(\ref{absorb3}). The absorption rate
\begin{math} \Gamma  \end{math} is completely determined by the
viscosity \begin{math} \eta  \end{math} via
\begin{equation}
\Gamma =\left(\frac{16\pi G}{c^2}\right)\eta
\label{absorb4}
\end{equation}
as has been proved in previous work\cite{viscosity}.

For a finite size medium, such as a gravitational wave antenna,
the rigorously exact result for a graviton to be absorbed at temperature
\begin{math} T \end{math} is determined by the
non-local viscosity. With \begin{math} \beta =(\hbar /k_BT) \end{math},
the microscopic Green-Kubo formula for viscosity is
\begin{equation}
\eta_{ijkl}({\bf r},{\bf r}^\prime ,\zeta )=
\int_{0}^\infty e^{i\zeta t}\left\{ \frac{1}{\hbar }\int_0^\beta
\left<p_{kl}({\bf r}^\prime ,-i\lambda )p_{ij}({\bf r},t)\right>
d\lambda \right\}dt.
\label{absorb5}
\end{equation}
\par \noindent
\textbf{Theorem:}{
The (LSZ reduction) formula for the {\em total cross section}
for a graviton with polarization \begin{math} {\sf e} \end{math}
at frequency \begin{math} \omega =c|{\bf k}| \end{math}
to be absorbed by an antenna of volume \begin{math} \Omega \end{math}
is given by
\begin{equation}
\sigma (\omega )=\frac{16\pi G}{c^3}
{\Re}e\int_\Omega \int_\Omega e^{i{\bf k}\cdot ({\bf r}^\prime-{\bf r})}
\{e^*_{ij}\eta_{ijkl}({\bf r},{\bf r}^\prime ,\omega+i0^+ )e_{kl}\}
d^3{\bf r} d^3{\bf r}^\prime .
\label{absorb6}
\end{equation}}

There can be no difference of opinion as to whether
the quantum pressure fluctuations producing the viscosity in
Eq.(\ref{absorb5})
determines the total graviton cross section as in Eq.(\ref{absorb6}).
That the {\em electrons dominate the viscosity} \begin{math} \eta \end{math}
in low temperature metals has been experimentally well
established\cite{el_vis1}. Since the cross section for the absorption
of gravity waves is rigorously determined by the
viscosity response function, it is then evident that electrons dominate
over the nuclear motions with regard to gravity wave absorption.

\section{Conclusion\label{conc}}

The gravitational wave is absorbed by an antenna via the quantum fluctuations
in the pressure, which describe the viscous response function. In a metallic
antenna, the electronic motions control the pressure and thereby control the
viscosity. The electronic nature of viscous electronic damping of sound has been
experimentally verified. Since viscosity also damps gravitational waves, it
seem natural that electronic motions should play a key role for gravitational
wave detection.

To understand the importance of electronic-gravitational wave interactions
from a microscopic viewpoint, one must keep the correct virial theorem pressure
contributions for both the Coulomb energy and kinetic energy. These contributions
have been correctly computed in previously discussed viscosity sum rules\cite{us}.
The sum rules dictate that the full inclusion of the Coulomb interactions be
present in an adequate microscopic theory. An inadequate model of pressure matrix elements
consists of electrons and phonons as two ideal gas components weakly interacting with
one another. In such weakly coupled electron-phonon models, the electronic Coulomb
interactions are ignored, removing the strongly non-adiabatic matrix elements required
for proper strength of electron-hole virtual excitations.

Finally, if the Coulomb interactions are {\em properly} included, then the
electronic enhancement of the gravitational wave detection efficiency is
theoretically substantial.

\end{document}